\documentclass[a4paper]{article}

\usepackage[english]{babel}
\usepackage[utf8x]{inputenc}
\usepackage[T1]{fontenc}

\usepackage[a4paper,top=3cm,bottom=2cm,left=3cm,right=3cm,marginparwidth=1.75cm]{geometry}

\usepackage{setspace}
\usepackage{amssymb}

\usepackage{cite}

\usepackage{amsmath}
\usepackage{graphicx}
\usepackage[colorinlistoftodos]{todonotes}

\title{Two-dimensional plasmonic waveguides for nanolasing and four-wave mixing}

\author{Guangyuan Li$^{1,*}$, Stefano Palomba$^{2,3}$, and C. Martijn de Sterke$^{2,3}$}

\date{}
\begin{document}
\maketitle

\begin{spacing}{2.0}

\noindent \large$^1$Shenzhen Institutes of Advanced Technology, Chinese Academy of Sciences, Shenzhen 518055, Guangdong Province, China

\noindent  $^2$Institute of Photonics and Optical Science (IPOS), School of Physics, The University of Sydney, NSW 2006, Australia

\noindent  $^3$The University of Sydney Nano Institute, The University of Sydney, NSW 2006, Australia

%\noindent $^\dagger$ These authors contributed equally.

\noindent *Corresponding author: gy.li@siat.ac.cn

\end{spacing}

\begin{abstract}
Plasmonic waveguides are an essential element of nanoscale coherent sources, including nanolasers and four-wave mixing (FWM) devices. Here we report how the design of the plasmonic waveguide needs to be guided by the ultimate application. This contrasts with traditional approaches in which the waveguide is considered in isolation. We find that hybrid plasmonic waveguides, with a nonlinear material sandwiched between the metal substrate and a high-index layer, are best suited for FWM applications, whereas metallic wedges are preferred in nanolasers. We also find that in plasmonic nanolasers high-index buffer layers perform better than more traditional low-index buffers.
\end{abstract}

\section{Introduction}
Waveguide-based nanoscale coherent light sources are essential to nanophotonic circuits in diverse applications including high-speed on-chip interconnects \cite{Miller2009InterConnect}, quantum information processing \cite{Lukin2014NanoPhotonQuantum}, chemical and biomedical sensing \cite{Eggleton2007optofluidics}. Plasmonic waveguides are key elements of such devices since they allow subwavelength light confinement, and hence greatly enhanced light-matter interactions \cite{Stockman2003SPASER,Zayats2012NLPlasReview}. We consider two different classes of plasmonic-waveguide based coherent nanoscale light sources: nanolasers \cite{XZhang2013PlaslaserRev,Gwo2016PlasNanoLaserRev,Odom2017PLrev}, and nonlinear devices for frequency conversion, specifically degenerate four-wave mixing (DFWM) \cite{Diaz2016ExpHPPNL,Raschke2016FWMimaging,Oulton2017DFWM_MDM}.

Although requirements on plasmonic waveguides for nanolasing and for DFWM applications are distinct, and are also different from those for pure waveguiding applications, plasmonic waveguides are traditionally characterized by their attenuation and by their mode confinement, independently of the ultimate application \cite{Oulton2008PlasWGrev}. For example, after comparing four typical plasmonic waveguides, Oulton {\sl et al.} found that hybrid plasmonic metal/low-index insulator/high-index semiconductor/air (known as MISA) configurations, exhibit low attenuation loss and high mode confinement \cite{Oulton2008PlasWGrev}. Subsequently, MISA-based devices have been widely used in plasmonic nanolasers \cite{Oulton2009PlasonLaser,Ma2011RTplasmonLaser,Gwo2012ExpiAgPlasmonLaser,Oulton2014UltrafastLaser,Zhang2014UVplasNanoLaser,Oulton2016PerovskitePlasmonLaser}. Encouraged by the attractive waveguiding characteristics, as well as the successful applications in nanolasers, the use of hybrid plasmonic waveguides has been extended to nonlinear applications \cite{Zhang2013HPPNL,Pitilakis2013NLFOMPWG,Diaz2016ExpHPPNL,Diaz2016TheoHPPNL}, but without a rigorous comparison to other types of plasmonic waveguides.

%Over the years, new plasmonic waveguide structures have been proposed and characterized with attenuation loss versus mode confinement [***], although the potential applications for nanolasing or DFWM applications, rather than waveguiding applications.

We recently developed a comprehensive theory for the design and analysis of plasmonic waveguides for nanolasing, and applied this theory, after a slight modification, to the waveguide design for plasmonic FWM devices as well \cite{Li2018Analogies}. Our exploration based on one-dimensional (1D) plasmonic waveguide geometries not only provides deep understanding of currently used designs, but also led to potentially superior structures, with the potential to address longstanding challenges in plasmonic nanolasers. Although one- and two-dimensional plasmonic waveguides share the same physics, edge effects introduced by additional dimension may change the modal properties of plasmonic waveguides \cite{Bozhevolnyi2013PlasWGrev} and accordingly the nanolasing or DFWM performance.

\begin{figure}[!hbt]
\centering
\includegraphics[width=\linewidth]{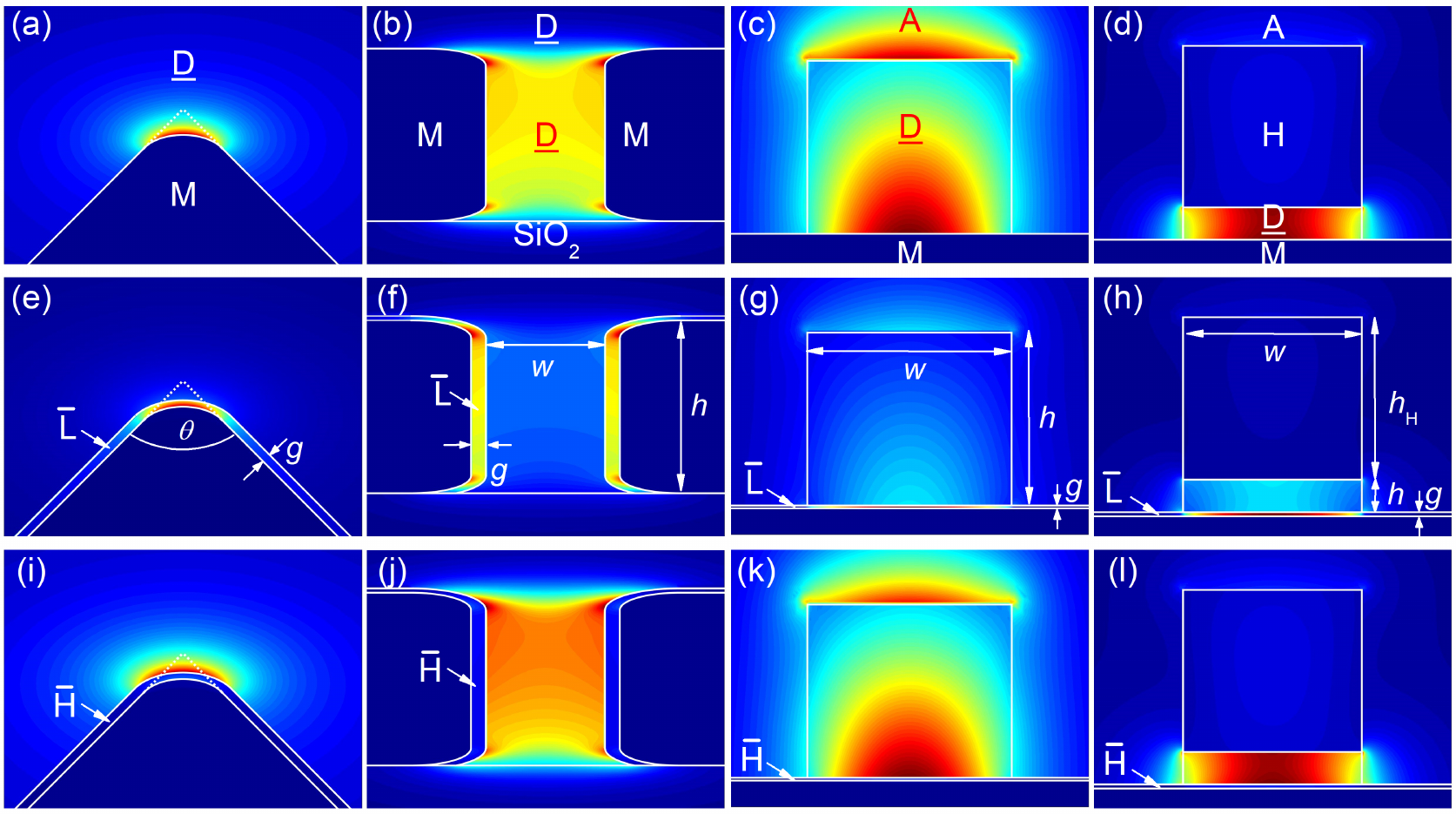}
\caption{Square modulus of the electric field $|\mathbf{e}|^2$ of the fundamental plasmonic mode for four 2D plasmonic waveguides (top panel) and their variations with a low-index (``$\bar{\rm L}$'', middle panel) and high-index (``$\bar{\rm H}$'', bottom panel) buffer layers of thickness $g=5$ nm. (a) M\underline{D} wedge waveguide with $\theta=60^\circ$; (b) M\underline{D}M slot waveguide with $h=300$\,nm and $w=20$\,nm; (c) M\underline{D}A DLSPP waveguide with $w=300$\,nm and $h=400$\,nm; and (d) M\underline{D}HA hybrid plasmonic waveguide with $h_{\rm H}=200$\,nm, $w=300$\,nm and $h=20$\,nm. All the structures are outlined by white curves. Parameters used: for $\lambda=1,550~{\rm nm}$, $n_{\rm M} = 0.1450 + 11.4125i$ (Ag), $n_0=1.8$ (DDMEBT) for the ``\underline{D}'' layer, $n_{\rm A}=1.0$ (Air), $n_{\rm H}=3.5$ (Si), and for the buffer layer $n_{\bar{\rm L}}=1.38$ (MgF$_2$) or $n_{\bar{\rm H}}=3.0$ (AlAs). }
\label{fig:struc}
\end{figure}

Here we characterize four realistic 2D plasmonic waveguides for nanolasing and DFWM applications. The four structures, illustrated in Figure~\ref{fig:struc}(a)--(d), are: a metallic wedge waveguide, a metal-slot waveguide, a dielectric loaded surface plasmon polariton (DLSPP) waveguide, and a hybrid plasmonic waveguide.

The outline of this paper is as follows. In Section~\ref{sec:compareNL} we compare the nonlinear contributions due to the metal with that due to a nonlinear dielectric, and show that the nonlinearity of the metal can be ignored in these DFWM devices. In Section~\ref{sec:measures} we then briefly introduce the theory we previously developed \cite{Li2018Analogies}, highlighting the characteristic measures of plasmonic waveguides for nanolasing and DFWM applications. The characteristics of plasmonic waveguides for use in nanolasers and DFWM devices can then be compared and understood in Section \ref{sec:comp}, allowing us to identify the desirable plasmonic waveguides for these applications. Section \ref{sec:buffer} includes a discussion of the effects of the refractive index of buffer layers that are used in nanolasers to avoid quenching.

%%%%%%%%%%%%%%%%%%%%%%%%%%%%%%%%%%%%%%%%%%%%%%%%%%%%%%%%%%%%%%%%%%%%%%%%%%%%%%%%%%%%%%%%%%%%%%%%%%
%%%%%%%%%%%%%%%%%%%%%%%%%%%%%%%%%%%%%%%%%%%%%%%%%%%%%%%%%%%%%%%%%%%%%%%%%%%%%%%%%%%%%%%%%%%%%%%%%%

\section{Nonlinear contributions from constituent materials}
\label{sec:compareNL}

In nonlinear plasmonic waveguides all constituent materials contribute to some degree to the overall nonlinearity. However, since our theory \cite{Li2018Analogies} assumes that the contribution from a highly nonlinear dielectric dominates, we need to validate this assumption before we apply this theory to a specific plasmonic waveguide structure. This is especially important since recent studies showed that the third-order nonlinear susceptibility $\chi^{(3)}$ of metals can be large at optical wavelengths \cite{Marini2013AuNL,Boyd2014AuNL,Qian2016AuKai3Quantum}.

To carry out this comparison we follow the approach by Baron {\sl et al.} for surface plasmon polaritons \cite{Baron2015SPPMD}, which we recently extended to more complicated 1D geometries \cite{Li2018Flimit}. For simplicity, we take only one dielectric to be highly nonlinear, and neglect nonlinear contributions from other dielectrics in the plasmonic waveguide. The ratio of the nonlinear contribution from the highly nonlinear dielectric and that from metal can be written as \cite{Li2018Flimit}
%%%%%%%%%%%%%%%%%%%%%%%%%%%%%%%%%%%%%%%%%%%%%%%%%%%%%%%%%%%%%%%%%%%%%%%%%%%%%%%%%%%%%%%%%%%%%%%%%%
\begin{eqnarray}
r_{\rm \underline{D}2M} \approx \left|\frac{\chi^{(3)}_{\rm \underline{D}}}{\chi^{(3)}_{\rm M}} \right|\frac{\int_{\rm \underline{D}} |\mathbf{e}|^4 \mathrm{d}A}{\int_{\rm M} |\mathbf{e}|^4 \mathrm{d}A} \,,
\label{eq:Ratio}
\end{eqnarray}
%%%%%%%%%%%%%%%%%%%%%%%%%%%%%%%%%%%%%%%%%%%%%%%%%%%%%%%%%%%%%%%%%%%%%%%%%%%%%%%%%%%%%%%%%%%%%%%%%%
where we used $(2|\mathbf{e}|^4 + |\mathbf{e}^2|^2)/3\approx |\mathbf{e}|^4$, $\mathbf{e}$ is the modal electric field, $\chi^{(3)}$ is the third-order nonlinear susceptibility, and ``\underline{D}'' and ``M'' refer to the highly nonlinear dielectric and the metal, respectively. The integrals are taken over the respective media.

The four 2D plasmonic waveguides we consider are shown in Figures~\ref{fig:struc}(a)--(d), and are denoted as ``M\underline{D}'' wedge, ``M\underline{D}M'' slot, ``M\underline{D}A'' DLSPP, and ``M\underline{D}HA'' hybrid plasmonic waveguide, respectively. These denotations indicate the type of layers encountered in the direction perpendicular to the metal surface. In addition to ``M'' and ``\underline{D}'' defined above, ``A'' means air, and ``H'' is a dielectric that has higher refractive index than ``D''. For all these waveguides, we consider only the fundamental plasmonic mode. For the ``M\underline{D}'' wedge, we take the radius of the metallic curvature to be $20\,$nm \cite{Kress2015Wedge}. %More complicate plasmonic waveguides can be constructed by adding more layers or by combining two structures, according to our 1D explorations \cite{Li2018Analogies}, or by changing the geometric shapes.\marginpar{Clark, it is not clear why this sentence is here--I think it needs a punch line.}

Throughout this work, we consider $\lambda= 1.550~{\rm\mu m}$, and take metal to be silver with linear refractive index $n_{\rm M} = 0.1450 + 11.4125i$ \cite{JC1972NK}, and third-order nonlinear susceptibility $\left|\chi^{(3)}_{\rm M} \right|=2.8 \times 10^{-19} \,{\rm m}^2 / {\rm V}^{2}$ \cite{Boyd_textbook_NLO}. We consider the highly nonlinear dielectric to be DDMEBT with linear refractive index $n_0=1.8$, nonlinear refractive index $n_2 = 1.7\times 10^{-17}\,{\rm m}^2/ {\rm W}$ \cite{Biaggio2008DDMEBTX3}, and thus $\chi^{(3)}_{\rm \underline{D}}=(4/3)c \varepsilon_0 n_0^2 n_2 =1.95\times 10^{-19} \,{\rm m}^2/ {\rm V}^{2}$. The field profiles of the fundamental modes are calculated using the finite element method as implemented in COMSOL, and are shown in Figure~\ref{fig:struc}.

%%%%%%%%%%%%%%%%%%%%%%%%%%%%%%%%%%%%%%%%%%%%%%%%%%%%%%%%%%%%%%%%%%%%%%%%%%%%%%%%%%%%%%%%%%%%%%%%%%
\begin{figure}[!hbt]
\centering
\includegraphics[width=\linewidth]{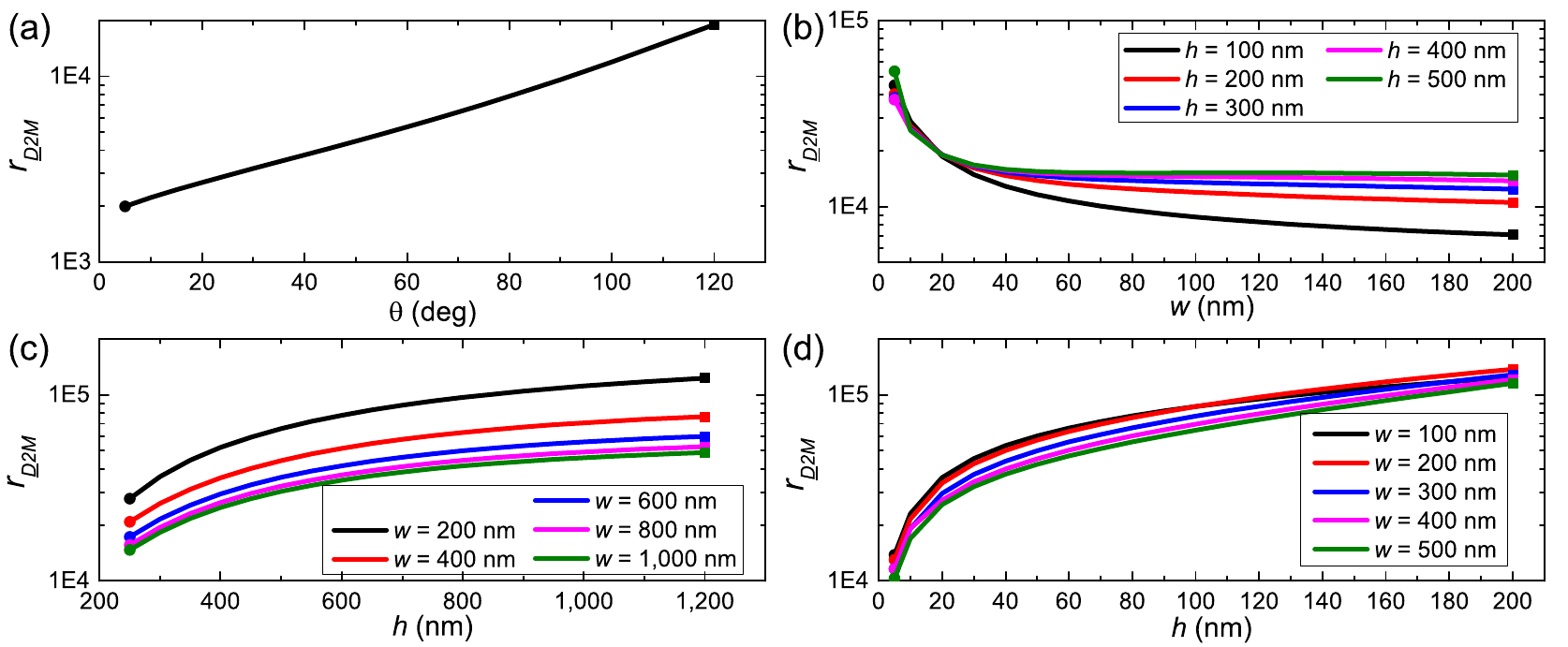}
\caption{Nonlinear contribution ratios $r_{\rm \underline{D}2M}$ for (a) M\underline{D} wedges, (b) M\underline{D}M slot waveguides, (c) M\underline{D}A DLSPP waveguides, and (d) M\underline{D}HA hybrid plasmonic waveguides with $h_{\rm H}=200$\,nm.}
\label{fig:NLratio}
\end{figure}
%%%%%%%%%%%%%%%%%%%%%%%%%%%%%%%%%%%%%%%%%%%%%%%%%%%%%%%%%%%%%%%%%%%%%%%%%%%%%%%%%%%%%%%%%%%%%%%%%%

Figure \ref{fig:NLratio} shows that for all the four plasmonic waveguides under study, $r_{\rm \underline{D}2M}\geq 2\times10^3$. This means that the nonlinear contribution from metal is negligible compared to that from the highly nonlinear dielectric. This is because although $\chi^{(3)}_{\rm \underline{D}} < \left|\chi^{(3)}_{\rm Ag} \right|$, we find $\int_{\rm \underline{D}} |\mathbf{e}|^4 \mathrm{d}A \gg \int_{\rm M} |\mathbf{e}|^4 \mathrm{d}A$ since the electric field is mainly confined to the highly nonlinear dielectric region (see Figure~\ref{fig:struc}(a)--(d)). This conclusion does not change even if silver is replaced by gold, which, according to a recent report \cite{Qian2016AuKai3Quantum} has a nonlinear susceptibility as large as $\left|\chi^{(3)}_{\rm Au} \right|=9.1\times 10^{-19} \,{\rm m}^2/ {\rm V}^{2}$.

%We note that the ratios $r_{\rm \underline{D}2M}$ in Figure~\ref{fig:NLratio} are much larger than those obtained by Nielsen {\sl et al.} \cite{Oulton2017DFWM_MDM}. This is because here we derive $r_{\rm \underline{D}2M}$ using the fully vectorial nonlinear coefficient $\gamma$ \cite{Afshar2009NLKerr,Li2017Gamma}, whereas the expression used by Nielsen {\sl et al.} is not rigorous, and overestimates the nonlinear contributions of metal. This highlights the importance of adopting rigorous expression for $\gamma$.

\section{Characteristic measures for plasmonic nanolasers and DFWM devices}
\label{sec:measures}

The theory we developed \cite{Li2018Analogies} shows that plasmonic waveguides used as nanolasers or as DFWM devices can be measured using a set of two characteristic dimensionless parameters. Surprisingly, these two sets of parameters are very similar, and each provides clear physical insights. We discuss these now.

The first characteristic measure for plasmonic waveguides that applies to both devices, is the effective area $A_{\rm eff}$ normalized by the diffraction-limited area $A_0 \equiv (\lambda/(2n_0))^2$, {\sl i.e.}, $A_{\rm eff}/A_0$. Here $A_{\rm eff}$ is defined as \cite{Li2018Analogies}
%%%%%%%%%%%%%%%%%%%%%%%%%%%%%%%%%%%%%%%%%%%%%%%%%%%%%%%%%%%%%%%%%%%%%%%%%%%%%%%%%%%%%%%%%%%%%%%%%%
\begin{eqnarray}
\label{eq:Aeff}
A_{\rm eff} \equiv \frac{ \int_\infty (\mathbf{e} \times \mathbf{h}^*) \cdot\hat{z} \mathrm{d}A} {{\rm max}\{c \varepsilon_0 n_0 |\mathbf{e}|^2\}_{\rm \underline{D}}}\,,
\end{eqnarray}
%%%%%%%%%%%%%%%%%%%%%%%%%%%%%%%%%%%%%%%%%%%%%%%%%%%%%%%%%%%%%%%%%%%%%%%%%%%%%%%%%%%%%%%%%%%%%%%%%%
where $\{\mathbf{e}, \mathbf{h}\}$ are the modal fields, ${\rm max}\{\cdot\}_{\rm \underline{D}}$ indicates the maximum over the active medium which has linear refractive index $n_0$. The active medium \underline{D} denotes the gain medium in nanolasers, and the highly nonlinear layer in DFWM devices.

In nanolasers, $A_{\rm eff}$ measures the maximum Purcell factor of a waveguide mode \cite{Sorger2011HPPEnhance}, $F_{\rm m,max}$, through $F_{\rm m,max}= (3/\pi) A_0/A_{\rm eff}$, whereas in DFWM devices, $A_{\rm eff}$ measures the driving power $P_{0,{\rm max}}$ for which the nonlinear refractive index change of the highly nonlinear material saturates and reaches its maximum value $\Delta n_{\rm max}$, via $P_{0,{\rm max}}=A_{\rm eff} I_{\rm bulk,max}$. Here $I_{\rm bulk,max}$ is the maximum allowed intensity under plane wave illumination before optical damage occurs.

The second dimensionless measure for plasmonic waveguides in a nanolaser is
%%%%%%%%%%%%%%%%%%%%%%%%%%%%%%%%%%%%%%%%%%%%%%%%%%%%%%%%%%%%%%%%%%%%%%%%%%%%%%%%%%%%%%%%%%%%%%%%%%
\begin{eqnarray}
\label{eq:Gth}
k_0/g_{\rm th} = \frac{\int_{\rm \underline{D}} n_0^2 |\mathbf{e}|^2 \mathrm{d}A} {n_0 \int_{\rm M} \varepsilon''_{\rm r,M} |\mathbf{e}|^2 \mathrm{d}A} \,,
\end{eqnarray}
%%%%%%%%%%%%%%%%%%%%%%%%%%%%%%%%%%%%%%%%%%%%%%%%%%%%%%%%%%%%%%%%%%%%%%%%%%%%%%%%%%%%%%%%%%%%%%%%%%
where $g_{\rm th}$ is the threshold gain, {\sl i.e.}, the material gain needed to compensate the modal loss and to achieve lasing. Further, $k_0=2\pi/\lambda$, and $\varepsilon_{\rm r,M}=\varepsilon'_{\rm r,M}+i\varepsilon''_{\rm r,M}$ is the relative permittivity of metal.

The second dimensionless measure for plasmonic waveguide in a DFWM device is
%%%%%%%%%%%%%%%%%%%%%%%%%%%%%%%%%%%%%%%%%%%%%%%%%%%%%%%%%%%%%%%%%%%%%%%%%%%%%%%%%%%%%%%%%%%%%%%%%%
\begin{eqnarray}
\label{eq:FOM}
{\cal F}/{\Delta n_{\rm max}} =  \frac{\int_{\rm \underline{D}} n_0^2 |\mathbf{e}|^2 \cdot U \mathrm{d}A} {n_0 \int_{\rm M} \varepsilon''_{\rm r,M} |\mathbf{e}|^2 \mathrm{d}A} \,,
\end{eqnarray}
%%%%%%%%%%%%%%%%%%%%%%%%%%%%%%%%%%%%%%%%%%%%%%%%%%%%%%%%%%%%%%%%%%%%%%%%%%%%%%%%%%%%%%%%%%%%%%%%%%
where ${\cal F}$ is the Figure of Merit, and $U\equiv |\mathbf{e}|^2 / {\rm max} \{|\mathbf{e}|^2 \}$ expresses the field uniformity in the highly nonlinear medium.

A low-threshold waveguide-based nanolaser requires (i) enhanced spontaneous emission that is efficiently coupled into the guided mode, which is equivalent to requiring the mode to have a high Purcell factor in order to make full use of the gain medium; and (ii) a low threshold gain $g_{\rm th}$. Optimizing a nanoscale DFWM device requires (i) that the maximum nonlinearity of the material, {\sl i.e.}, the maximum achievable nonlinear index change is reached at a low driving power, $P_{0,{\rm max}}$; and (ii) a large Figure of Merit ${\cal F}$ in order to achieve a high conversion efficiency. We showed earlier that the maximum achievable conversion efficiency is $\eta_{\rm max}=4 {\cal F}^2 /27$ \cite{Li2016FOMKerrPW}, which is defined as the ratio of the output idler power to the input signal power. Requirements (i) for both types of devices are equivalent to a small effective area $A_{\rm eff}/A_0$, whereas requirements (ii) for both devices correspond to large $k_0/g_{\rm th}$ and large ${\cal F}/{\Delta n_{\rm max}}$.

We can further write
%%%%%%%%%%%%%%%%%%%%%%%%%%%%%%%%%%%%%%%%%%%%%%%%%%%%%%%%%%%%%%%%%%%%%%%%%%%%%%%%%%%%%%%%%%%%%%%%%%
\begin{eqnarray}
k_0/g_{\rm th} &= \Gamma_{\rm G} \cdot (k_0 L_{\rm att})\,,\\
{\cal F}/{\Delta n_{\rm max}} &= ({\rm EFF}_{\rm NL}/f_{\ell}) \cdot (k_0 L_{\rm att}) \,.
\end{eqnarray}\label{eq:FOMgth}
%%%%%%%%%%%%%%%%%%%%%%%%%%%%%%%%%%%%%%%%%%%%%%%%%%%%%%%%%%%%%%%%%%%%%%%%%%%%%%%%%%%%%%%%%%%%%%%%%%
Here the gain confinement factor $\Gamma_{\rm G}$ \cite{Lipson2008GianConf} and the nonlinear effectiveness $ {\rm EFF}_{\rm NL}$ \cite{Li2018Flimit} can be written as
%%%%%%%%%%%%%%%%%%%%%%%%%%%%%%%%%%%%%%%%%%%%%%%%%%%%%%%%%%%%%%%%%%%%%%%%%%%%%%%%%%%%%%%%%%%%%%%%%%
\begin{equation}
\label{eq:Gconf}
\Gamma_{\rm G} = \frac{Z_0\int_{\rm \underline{D}} n_0^2 |\mathbf{e}|^2 \mathrm{d}A} {n_0 \int_\infty (\mathbf{e} \times \mathbf{h}^*) \cdot \hat{z}  \mathrm{d}A}; \hspace{10mm}
{{\rm EFF}_{\rm NL}\over f_{\ell}} = \frac{Z_0\int_{\rm \underline{D}} n_0^2 |\mathbf{e}|^2 \cdot U \mathrm{d}A} {n_0 \int_\infty (\mathbf{e} \times \mathbf{h}^*) \cdot \hat{z}  \mathrm{d}A} \,,
\end{equation}
%%%%%%%%%%%%%%%%%%%%%%%%%%%%%%%%%%%%%%%%%%%%%%%%%%%%%%%%%%%%%%%%%%%%%%%%%%%%%%%%%%%%%%%%%%%%%%%%%%
where $Z_0$ is the vacuum impedance, and $f_{\ell}= 2/(3\ln{(3)})$ accounts for modal loss \cite{Li2016FOMKerrPW}.

Comparing Eqs~(\ref{eq:Gth}) and (\ref{eq:FOM}), or Eqs~(\ref{eq:Gconf}), we find that the only difference between $k_0/g_{\rm th}$ or $\Gamma_{\rm G}$ for plasmonic nanolasers and ${\cal F}/{\Delta n_{\rm max}}$ or ${\rm EFF}_{\rm NL}/f_{\ell}$ for plasmonic DFWM devices is that the latter has an additional factor $U$ in the integral kernel of the numerator. This factor highlights the importance of field uniformity in the active medium for nonlinear DFWM applications.

Compared with the parameters $A_{\rm eff}/A_0$ and $L_{\rm att}$ for characterizing plasmonic waveguides for pure waveguiding applications, it is clear that the measures for nanolasing and DFWM, $A_{\rm eff}/A_0$ and $k_0/g_{\rm th}$ or ${\cal F}/{\Delta n_{\rm max}}$, also include $\Gamma_{\rm G}$ or ${\rm EFF}_{\rm NL}/f_{\ell}$, corresponding to the overlap between the mode field and the gain or nonlinear medium. Therefore, plasmonic waveguides that are optimized for waveguiding applications may be suboptimal for nanolasing or DFWM. 

\section{Characteristics for nanolasing and DFWM}\label{sec:comp}

Figure \ref{fig:FOMnormk02gth} compares the nanolasing and DFWM characteristics of the four 2D plasmonic waveguides from Figure~\ref{fig:struc}(a)--(d).  We now discuss these in turn. At this stage we neglect the quenching effects in plasmonic nanolasers; we shall consider these in Section \ref{sec:buffer}.

\subsection{M\underline{D} wedge waveguide}
\begin{figure}[!hbt]
\centering
\includegraphics[width=\linewidth]{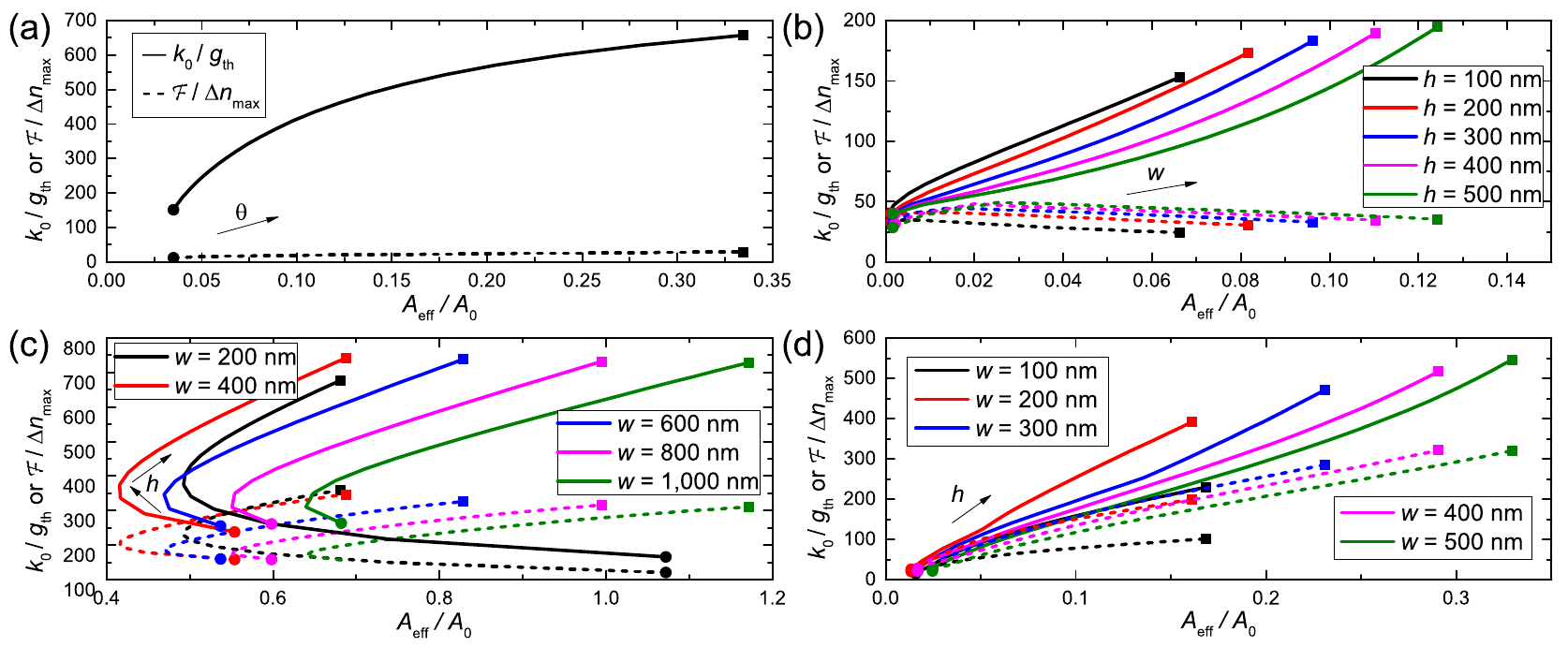}
\caption{Nanolasing ($k_0/g_{\rm th}$ in solid lines) and DFWM (${\cal F}/{\Delta n_{\rm max}}$ in dashed lines) characteristics of (a) M\underline{D} wedges with $\theta=[5^\circ, 120^\circ]$, (b) M\underline{D}M slot waveguides with $w=[5,200]$\,nm, (c) M\underline{D}A  DLSPP waveguides with $h=[250,1200]$\,nm, and (d) hybrid plasmonic M\underline{D}HA  waveguides with $h_{\rm H}=200$\,nm and $h=[5,200]$\,nm. The starting and ending points of the variables are indicated by a circle and a rectangle, respectively.}
\label{fig:FOMnormk02gth}
\end{figure}

Figure \ref{fig:FOMnormk02gth}(a) shows that, as the wedge angle $\theta$ for M\underline{D} wedge waveguides increases from $5^\circ$ to $120^\circ$, $A_{\rm eff}/A_0$ increases from $0.035$ to $0.335$, thus remaining below the diffraction limit ($A_{\rm eff}/A_0<1$). Simultaneously, $k_0/g_{\rm th}$ increases from $150$ to $657$. In other words, the maximum modal Purcell factor $F_{\rm m,max}$ decreases from $27$ to $3$, while the gain threshold $g_{\rm th}$ decreases from $270\,{\rm cm}^{-1}$ to $62\,{\rm cm}^{-1}$. On the other hand, ${\cal F}/{\Delta n_{\rm max}}$ increases only slightly  and remains below $30$. Thus the maximum achievable DFWM conversion efficiency $\eta_{\rm max}$ would be low: $\eta_{\rm max}<(400/3)(\Delta n_{\rm max})^2 = 0.33\%$ for $\Delta n_{\rm max}=0.5\%$, making M\underline{D} wedge waveguides less suited for DFWM applications.
%%%%%%%%%%%%%%%%%%%%%%%%%%%%%%%%%%%%%%%%%%%%%%%%%%%%%%%%%%%%%%%%%%%%%%%%%%%%%%%%%%%%%%%%%%%%%%%%%%
%%%%%%%%%%%%%%%%%%%%%%%%%%%%%%%%%%%%%%%%%%%%%%%%%%%%%%%%%%%%%%%%%%%%%%%%%%%%%%%%%%%%%%%%%%%%%%%%%%
%%%%%%%%%%%%%%%%%%%%%%%%%%%%%%%%%%%%%%%%%%%%%%%%%%%%%%%%%%%%%%%%%%%%%%%%%%%%%%%%%%%%%%%%%%%%%%%%%%
\begin{figure}[!hbt]
\centering
\includegraphics[width=\linewidth]{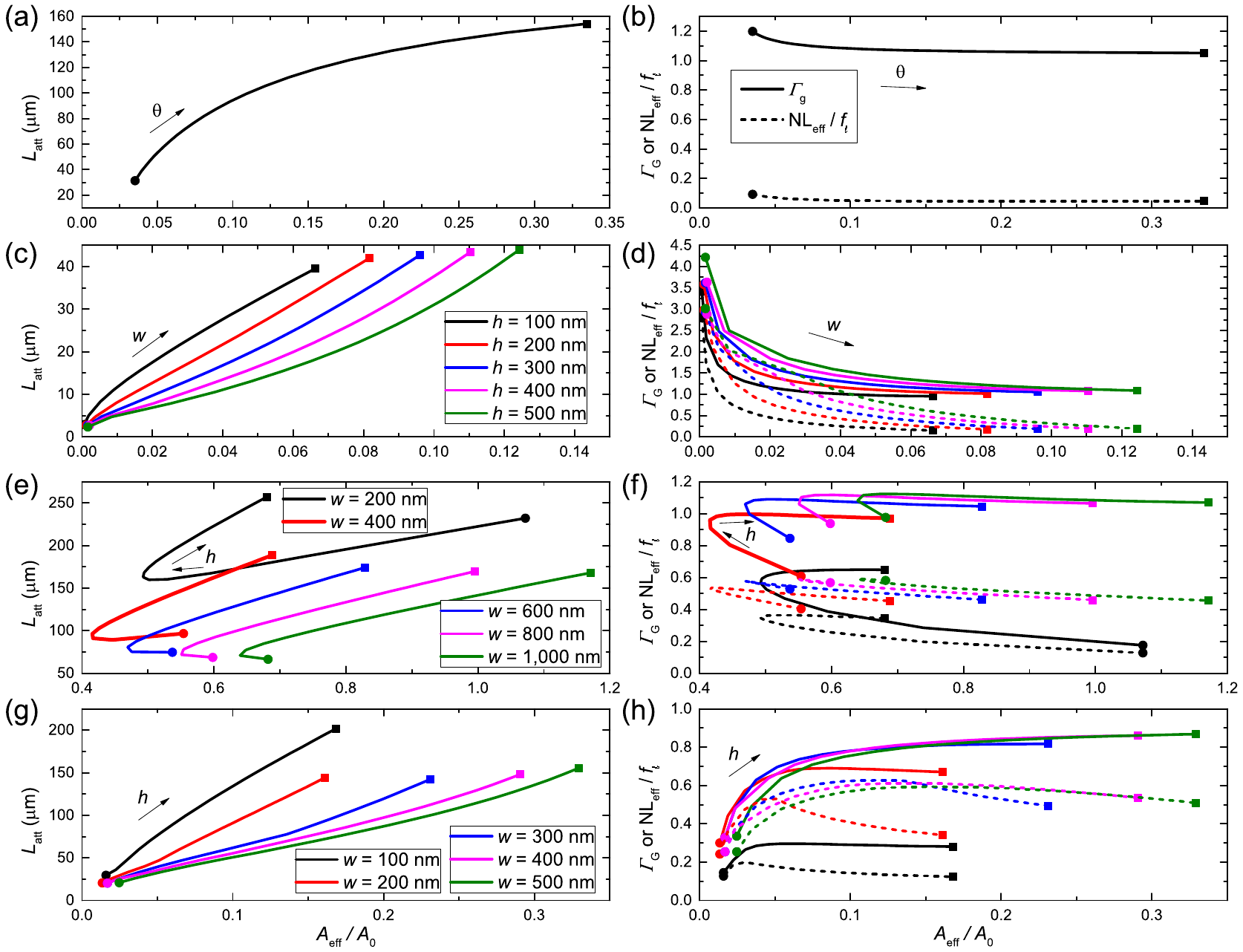}
\caption{Waveguiding and active (gain/nonlinear) characteristics of (a)(b) M\underline{D} wedges, (c)(d) M\underline{D}M slot waveguides, (e)(f) M\underline{D}A DLSPP waveguides, and (g)(h) hybrid plasmonic M\underline{D}HA waveguides. All the parameters are same as Figure~\ref{fig:FOMnormk02gth}.}
\label{fig:LattConfG}
\end{figure}
%%%%%%%%%%%%%%%%%%%%%%%%%%%%%%%%%%%%%%%%%%%%%%%%%%%%%%%%%%%%%%%%%%%%%%%%%%%%%%%%%%%%%%%%%%%%%%%%%%

The large difference between the nanolasing characteristics ($k_0/g_{\rm th}$ versus $A_{\rm eff}/A_0$) and the DFWM characteristics (${\cal F}/{\Delta n_{\rm max}}$ versus $A_{\rm eff}/A_0$) can be explained by (\ref{eq:FOMgth}). Although the M\underline{D} wedge waveguide combines long propagation length $L_{\rm att}$ and small effective area $A_{\rm eff}/A_0$, as shown by Figure~\ref{fig:LattConfG}(a), making it particularly attractive for waveguiding applications, the gain confinement factor $\Gamma_{\rm G}$ is large whereas the nonlinear effectiveness ${\rm EFF}_{\rm NL}$ is very low, as shown by Figure~\ref{fig:LattConfG}(b). This difference comes from the $U$ factor, which is important for M\underline{D} wedge waveguides because of the extremely nonuniform modal electric field of the mode (see Figure~\ref{fig:struc}(a)).

\subsection{M\underline{D}M slot waveguide}
Figure \ref{fig:FOMnormk02gth}(b) compares the nanolasing and DFWM characteristics of M\underline{D}M slot waveguides. $A_{\rm eff}/A_0$ can reach the deep subdiffraction regime ($\sim 1\times10^{-3}$) for narrow slot widths of $w=10$ nm, making the M\underline{D}M slot waveguide particularly attractive for achieving ultra-high modal Purcell factor in nanolasing applications or ultra-low driving power in nonlinear DFWM applications. For narrow slots $k_0/g_{\rm th}$ and ${\cal F}/{\Delta n_{\rm max}}$ have close values. %thanks to the uniform electric field in the slot filled with active medium, as shown by Figure~\ref{fig:struc}(b).
As the slot becomes wider, $k_0/g_{\rm th}$ increases whereas ${\cal F}/{\Delta n_{\rm max}}$ decreases in general.

These characteristics can also be understood using (\ref{eq:FOMgth}), where $L_{\rm att}$, $\Gamma_{\rm G}$ and ${\rm EFF}_{\rm NL}/f_{\ell}$ are shown in Figures~\ref{fig:LattConfG}(c)(d). As the slot width increases, more electric energy is confined to the slot, resulting in less attenuation loss or longer $L_{\rm att}$. However, the gain confinement factor $\Gamma_{\rm G}$ and the nonlinear effectiveness ${\rm EFF}_{\rm NL}$ both decrease unexpectedly; and more strikingly, they can be much larger than unity for narrow slot widths, indicating that the modal gain/nonlinearity is much larger than the gain/nonlinearity of the bulk material. This behaviour arises from slow light effects. We have shown \cite{Li2018Flimit} that both $\Gamma_{\rm G}$ and ${\rm EFF}_{\rm NL}/f_{\ell}$ can be factorized in an intuitive way as the product of factors that express the effects of losses, of slow light, and of the electric energy overlap with the active medium,
%%%%%%%%%%%%%%%%%%%%%%%%%%%%%%%%%%%%%%%%%%%%%%%%%%%%%%%%%%%%%%%%%%%%%%%%%%%%%%%%%%%%%%%%%%%%%%%%%%
\begin{eqnarray}
k_0/g_{\rm th} &= S \cdot \overline{\rho}_{\rm G}\,,\\
{\rm EFF}_{\rm NL}/f_{\ell} &=  S \cdot \overline{\rho}_{\rm NL} \,.
\label{eq:SlowLight}
\end{eqnarray}
%%%%%%%%%%%%%%%%%%%%%%%%%%%%%%%%%%%%%%%%%%%%%%%%%%%%%%%%%%%%%%%%%%%%%%%%%%%%%%%%%%%%%%%%%%%%%%%%%%
Here $S\equiv v_{\rm pw}/v_{\rm e}$ is the slow-down factor, with $v_{\rm pw} \equiv c/n_0$ the velocity of a plane wave in bulk and $v_{\rm e}=P/W$ the energy velocity of the mode. Here the mode power $P=\frac{1}{2}\int_{\infty} (\mathbf{e} \times \mathbf{h}^*) \cdot \hat{z}\mathrm{d}A$, and the mode energy $W=\frac{1}{4}\int_{\infty} \left[\mu_0 |\mathbf{h}|^2 + \varepsilon_0 [\partial (\omega \varepsilon'_{\rm r})/\partial \omega ]_{\omega_0} |\mathbf{e}|^2 \right] \mathrm{d}A$. $\overline{\rho}_{\rm G}$ ($0\leq \overline{\rho}_{\rm G} \leq 1$) quantifies the fraction of electric energy in the gain material, and $\overline{\rho}_{\rm NL}$ ($0\leq \overline{\rho}_{\rm NL} \leq 1$) quantifies fraction of the electric energy ratio in the nonlinear material, weighted by the field uniformity ($|\mathbf{e}|^2/{\rm max}\{| \mathbf{e}|^2\}$). Here
%%%%%%%%%%%%%%%%%%%%%%%%%%%%%%%%%%%%%%%%%%%%%%%%%%%%%%%%%%%%%%%%%%%%%%%%%%%%%%%%%%%%%%%%%%%%%%%%%%
\begin{eqnarray}
\label{eq:WtEnergyRatio}
\overline{\rho}_{\rm G} &\approx \frac{\int_{\rm \underline{D}} n_0^2 |\mathbf{e}|^2  \mathrm{d}A }{ \int_{\infty} [\partial (\omega \varepsilon'_{\rm r})/\partial \omega ]_{\omega_0} |\mathbf{e}|^2 \mathrm{d}A }\,,\\
\label{eq:WtEnergyRatio2}
\overline{\rho}_{\rm NL} &\approx \frac{\int_{\rm \underline{D}} n_0^2 |\mathbf{e}|^2 \cdot U \mathrm{d}A }{ \int_{\infty} [\partial (\omega \varepsilon'_{\rm r})/\partial \omega ]_{\omega_0} |\mathbf{e}|^2 \mathrm{d}A }\,,
\end{eqnarray}
%%%%%%%%%%%%%%%%%%%%%%%%%%%%%%%%%%%%%%%%%%%%%%%%%%%%%%%%%%%%%%%%%%%%%%%%%%%%%%%%%%%%%%%%%%%%%%%%%%
where the approximations originate from $2W\approx \int_{\infty} [\partial (\omega \varepsilon'_{\rm r})/\partial \omega ]_{\omega_0} |\mathbf{e}|^2 \mathrm{d}A$. For metal, which is dispersive, $[\partial (\omega \varepsilon'_{\rm r,M})/\partial \omega ]_{\omega_0}=\varepsilon'_{\rm r,M} + \omega_0 (\partial \varepsilon'_{\rm r,M}/\partial \omega)_{\omega_0}$, whereas $[\partial (\omega \varepsilon'_{\rm r})/\partial \omega ]_{\omega_0} =n_0^2$ for the active material which is taken to be dispersionless. Equations~(\ref{eq:WtEnergyRatio}) and (\ref{eq:WtEnergyRatio2}) imply that a large $\overline{\rho}_{\rm G}$ requires strong electric energy confinement in the gain material, whereas a large $\overline{\rho}_{\rm NL}$ requires strong and uniform electric energy confinement in the nonlinear material. For M\underline{D}M slot waveguides with narrow slots, the slow-down factor $S$ is very large, while the electric energy is strongly and uniformly confined to the slot \cite{Li2018Flimit}, leading to larger-than-unity $\Gamma_{\rm G}$ and ${\rm EFF}_{\rm NL}/f_{\ell}$. As the slot width increases, $S$ strongly decreases, while $\overline{\rho}_{\rm G}$ and $\overline{\rho}_{\rm NL}$ changes slightly \cite{Li2018Flimit}, resulting in decreasing $\Gamma_{\rm G}$ and ${\rm EFF}_{\rm NL}/f_{\ell}$.

As the M\underline{D}M slot becomes deeper, {\sl i.e.}, as $h$ increases, $A_{\rm eff}/A_0$, $k_0/g_{\rm th}$ and ${\cal F}/{\Delta n_{\rm max}}$ all slightly increase. The argument is similar to that for varying $w$, and is thus not further discussed here.

\subsection{M\underline{D}A DLSPP waveguide}
Figure \ref{fig:FOMnormk02gth}(c) shows that for M\underline{D}A DLSPP waveguides $A_{\rm eff}/A_0$ first decreases and then increases as the dielectric height increases. This can be understood following Holmgaard and Bozhevolnyi \cite{Bozhevolnyi2007DLSPP}: decreasing the height of the \underline{D} layer squeezes the field inside the \underline{D} region, but this behaviour does not continue indefinitely. Thus $A_{\rm eff}$ decreases until reaching a minimum; decreasing the \underline{D} height further results in a rapid increase in the field outside the \underline{D} region, resulting in increasing $A_{\rm eff}$. Similarly, $L_{\rm att}$ first decreases and then increases as the dielectric height increases, as shown by Figure~\ref{fig:LattConfG}(e). However, because of the complicated compromise between $L_{\rm att}$ and $\Gamma_{\rm G}$ (or ${\rm EFF}_{\rm NL}/f_{\ell}$) in Figure~\ref{fig:LattConfG}(f), $k_0/g_{\rm th}$ (or ${\cal F}/{\Delta n_{\rm max}}$) increases monotonously with height.

For M\underline{D}A DLSPP waveguides, the distinct characteristics for waveguiding applications, {\sl i.e.}, $L_{\rm att}$ versus $A_{\rm eff}/A_0$, and for nanolasing or DFWM applications, {\sl i.e.}, $\Gamma_{\rm G}$ or ${\rm EFF}_{\rm NL}/f_{\ell}$ versus $A_{\rm eff}/A_0$, clearly illustrate that the plasmonic waveguide optimized for waveguiding applications may not be optimal for nanolasing or DFWM. For example, Figure~\ref{fig:LattConfG}(e) shows that M\underline{D}A DLSPP waveguides with $h=200$\,nm generally have better waveguiding performance than those with $h=400$\,nm, {\sl i.e.}, have longer propagation lengths under the same confinement. %, according to the criteria used by Oulton {\sl et al.} \cite{Oulton2008PlasWGrev}.
However, because the latter have much larger $\Gamma_{\rm G}$ or ${\rm EFF}_{\rm NL}/f_{\ell}$, the net effect is that M\underline{D}A waveguides with $h=400$\,nm generally have better nanolasing or DFWM performance.

\subsection{M\underline{D}HA hybrid plasmonic waveguide}
For M\underline{D}HA hybrid plasmonic waveguides, Figure~\ref{fig:FOMnormk02gth}(d) shows that $k_0/g_{\rm th}$, ${\cal F}/{\Delta n_{\rm max}}$ and $A_{\rm eff}/A_0$ all increase, with increasing height of the active medium $h$. This is because $L_{\rm att}$ strongly increases, whereas $\Gamma_{\rm G}$ first increases and then saturates, and ${\rm EFF}_{\rm NL}/f_{\ell}$ first increases and then drops, as shown by Figures~\ref{fig:LattConfG}(g) and \ref{fig:LattConfG}(h), respectively. The differences between $\Gamma_{\rm G}$ and ${\rm EFF}_{\rm NL}/f_{\ell}$, and thus between $k_0/g_{\rm th}$ and ${\cal F}/{\Delta n_{\rm max}}$ are relatively small. This is because even for large hight $h$, the electric field in the active medium is relatively uniform, as shown in Figure~\ref{fig:struc}(d).

Similar to M\underline{D}A DLSPP waveguides, the waveguiding characteristics and the nanolasing or DFWM characteristics for M\underline{D}HA waveguides may be quite different. For example, Figure~\ref{fig:LattConfG}(g) shows that M\underline{D}HA waveguides with $h=100$\,nm have better waveguiding performance but worse nanolasing or DFWM performance than those with $h=200$\,nm. The reason, again, is because the latter have much larger $\Gamma_{\rm G}$ or ${\rm EFF}_{\rm NL}/f_{\ell}$, as shown in Figure~\ref{fig:LattConfG}(h).

\subsection{Comparison of four plasmonic waveguides}
\begin{figure*}
\centering
\includegraphics[width=\linewidth]{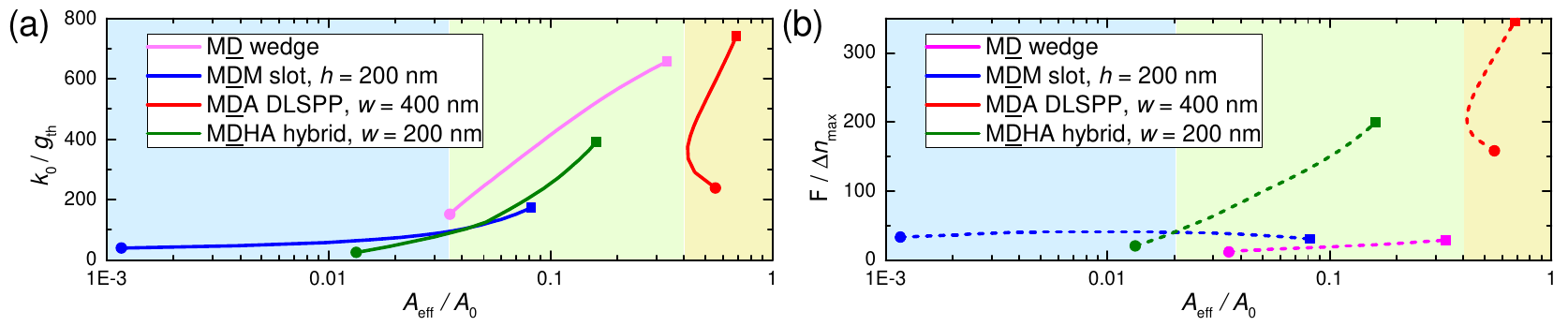}
\caption{Direct comparison of the (a) nanolasing and (b) DFWM characteristics of all four plasmonic waveguides under study. Waveguides M\underline{D}M slot with $h = 200$\,nm, M\underline{D}A DLSPP waveguide $w = 400$\,nm, M\underline{D}HA hybrid with $w = 200$\,nm, and the other parameters including the varying parameters are the same as Figure~\ref{fig:FOMnormk02gth}.}
\label{fig:Comparison}
\end{figure*}

Having understood the nanolasing and DFWM characteristics of all four plasmonic waveguide structures, we now summarize their performance in  Figure~\ref{fig:Comparison}. Depending on the effective area, we define the deep-, moderate- and near-subdiffraction regions, as indicated by blue, green, and yellow in Figure~\ref{fig:Comparison}.

Figure~\ref{fig:Comparison} shows that M\underline{D}M slot waveguides perform best in the deep-subdiffraction region since it combines the highest modal Purcell factor in nanolasing, and the lowest driving power in DFWM applications. %In other words, the M\underline{D}M slot waveguides are the best performing structures for both nanolasing and DFWM applications.
In the moderate-subdiffraction region, the M\underline{D} wedge waveguide has the best performance for use as a nanolaser since it combines high modal Purcell factor and low gain threshold (Figure~\ref{fig:Comparison}(a)), whereas the M\underline{D}HA hybrid waveguide performs best for use as a DFWM device by combining low driving power and high conversion efficiency (Figure~\ref{fig:Comparison}(b)). %This combination of high modal Purcell factor and low gain threshold makes the M\underline{D} wedge waveguide appealing for use as a plasmonic nanolaser or a plasmonic amplifier.
This is broadly consistent with the conclusion reached based on the waveguiding characteristics only \cite{Oulton2008PlasWGrev}. As a example, metallic wedge waveguides with integrated reflectors and precisely placed colloidal quantum dots have been used to achieve high-quality quantum emitters \cite{Kress2015Wedge}. In the near-subdiffraction region, or when we can tolerate a modest modal Purcell factor for nanolasers or high driving power for DFWM devices, the M\underline{D}A DLSPP performs best.

In our previous work on 1D plasmonic waveguides \cite{Li2018Analogies}, the 1D counterpart of the M\underline{D} wedge waveguide is the 1D M\underline{D} surface plasmon polariton (SPP) waveguide. Therefore, the pink curve in Figure~\ref{fig:Comparison} reduces to a point for its 1D counterpart  \cite{Li2018Analogies}. Therefore, in the moderate-subdiffraction region the best performing structure for both nanolasing and DFWM applications is the M\underline{D}HA hybrid plasmonic waveguide. In contrast, in the deep-, and the near-subdiffraction regions the wedge waveguide is the best performing 2D plasmonic waveguide configurations for both nanolasing and for DFWM, in line with our conclusions based on a 1D analysis \cite{Li2018Analogies}.

%%%%%%%%%%%%%%%%%%%%%%%%%%%%%%%%%%%%%%%%%%%%%%%%%%%%%%%%%%%%%%%%%%%%%%%%%%%%%%%%%%%%%%%%%%%%%%%%%%
%%%%%%%%%%%%%%%%%%%%%%%%%%%%%%%%%%%%%%%%%%%%%%%%%%%%%%%%%%%%%%%%%%%%%%%%%%%%%%%%%%%%%%%%%%%%%%%%%%

\section{Effect of buffer layers in plasmonic nanolasers}\label{sec:buffer}
For plasmonic nanolasers, a buffer layer needs to be included between the metal ``M'' and the gain material ``\underline{D}'' so as to avoid quenching. We recently showed that a high-index buffer outperforms a low-index buffer for 1D plasmonic waveguides, and illustrated the 2D extension of this with a specific M\underline{D}A 2D structure \cite{Li2018Analogies}. We now show that this conclusion also applies to the four 2D plasmonic waveguides studied in this work.

%%%%%%%%%%%%%%%%%%%%%%%%%%%%%%%%%%%%%%%%%%%%%%%%%%%%%%%%%%%%%%%%%%%%%%%%%%%%%%%%%%%%%%%%%%%%%%%%%%
\begin{figure*}
\centering
\includegraphics[width=\linewidth]{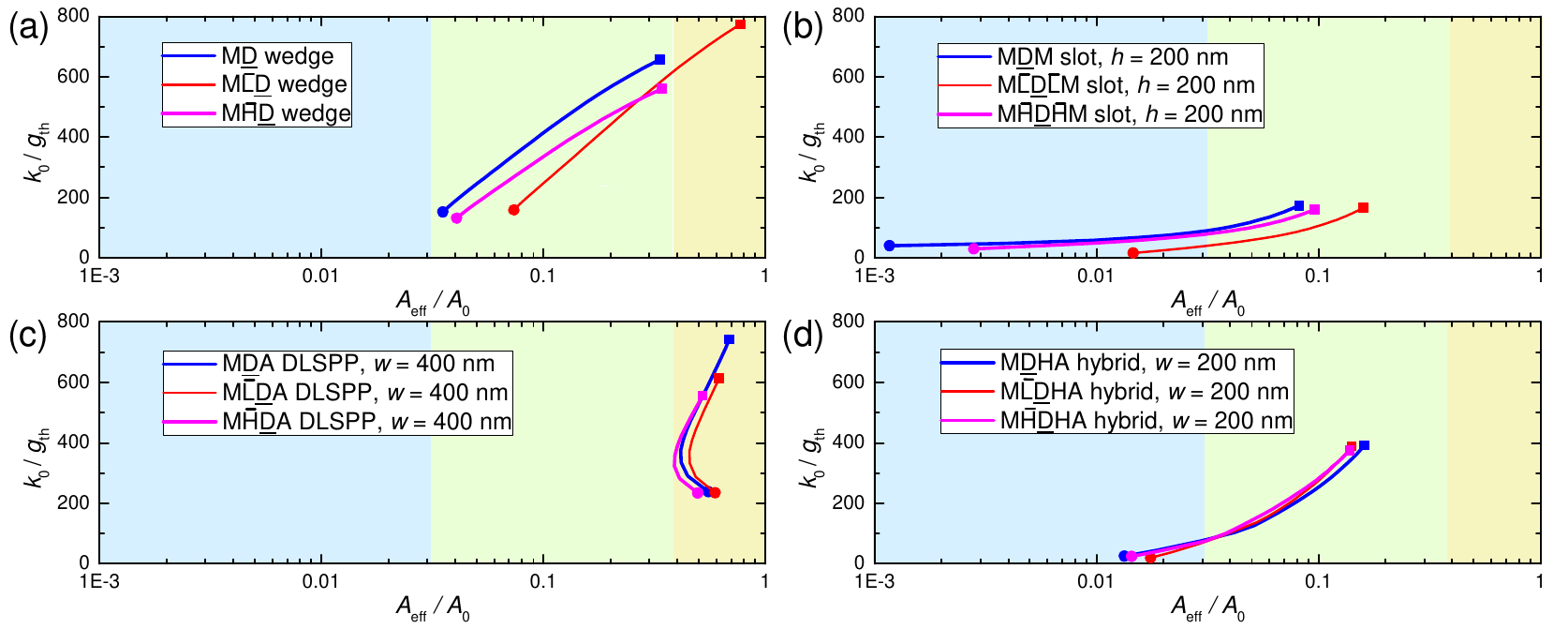}
\caption{Effects of low- (indicated by ``$\bar{\rm L}$'' with $n_{\bar{\rm L}}=1.38$) and high-index (``$\bar{\rm H}$'' with $n_{\bar{\rm H}}=3.0$) buffer layers (of thickness $g=5$ nm and sandwiched between ``M'' and ``\underline{D}'') on the nanolasing characteristics of all four plasmonic waveguides under study: (a) M\underline{D} wedges, (b) M\underline{D}M slot with $h = 200$\,nm, (c) M\underline{D}A DLSPP waveguide $w = 400$\,nm, and (d) M\underline{D}HA hybrid with $w = 200$\,nm. The other parameters, including the varying parameters, are the same as Figure~\ref{fig:FOMnormk02gth}. Structures without buffer layers, and with ``$\bar{\rm L}$'' and ``$\bar{\rm H}$'' buffers are indicated by blue, red and purple curves, respectively.}
\label{fig:buff}
\end{figure*}
%%%%%%%%%%%%%%%%%%%%%%%%%%%%%%%%%%%%%%%%%%%%%%%%%%%%%%%%%%%%%%%%%%%%%%%%%%%%%%%%%%%%%%%%%%%%%%%%%%

Figure \ref{fig:buff} shows that for all four 2D plasmonic waveguides the use of a high-index (indicated as ``$\bar{\rm H}$'' between ``M'' and ``\underline{D}'') buffer layer results in a smaller effective area than a low-index (``$\bar{\rm L}$'') buffer layer for the same lasing threshold. The thicknesses of all buffer layers, which can be deposited with atomic layer deposition or sputtering, are taken to be $g=5$ nm. In other words, in plasmonic nanolasers based on the four 2D plasmonic waveguides, a high-index buffer layer outperforms a low-index one. This conclusion is consistent with that for 1D configurations \cite{Li2018Analogies}. The reason is very similar: the use of a ``$\bar{\rm L}$'' buffer confines most of the electric field to the buffer layer (second row of Figure~\ref{fig:struc}), thus greatly decreasing the gain confinement factor, although the loss is also reduced. In contrast, ``$\bar{\rm H}$'' buffers confine most of the electric field to the gain ``\underline{D}'' medium (third row of Figure~\ref{fig:struc}), thus greatly increasing the gain confinement factor, which outweighs the slightly larger loss \cite{Li2018Analogies}.

\section{Conclusions}
In conclusions, we have investigated the nanolasing and DFWM characteristics of four 2D plasmonic waveguides. Our analysis shows that the M\underline{D}M slot waveguide is the best performing structure for achieving the highest modal Purcell factor in nanolasing or the lowest driving power in DFWM applications. The M\underline{D} wedge waveguide has the best performance for use as a nanolaser by combimning high model Purcell factor and low gain threshold, whereas the M\underline{D}HA hybrid waveguide performs best for use as a DFWM device for combing low driving power and high conversion efficiency. We also show that for plasmonic nanolasers that require a buffer layer between the metal and the gain medium, high-index buffers outperform conventional low-index buffers. We expect this work will advance the design and understanding of plasmonic waveguides that are used in nanolasers and nanoscale DFWM devices.

\section*{Acknowledgments}
This work was supported by the School of Physics, the University of Sydney, by the Shenzhen Research Foundation (Grant No. JCYJ20180507182444250), and by the State Key Laboratory of Advanced Optical Communication Systems and Networks, China (No. 2019GZKF2).
%\end{acknowledgments}

\end{document}